\documentclass[aps,
footinbib,
superscriptaddress,
showpacs,twocolumn,
prl]{revtex4-1}

\usepackage{amsmath}
\usepackage{amsfonts}  
\usepackage{graphicx}
\usepackage{array}
\usepackage{hyperref}
\usepackage[usenames]{color}

\usepackage{braket}

\renewcommand{\v}[1]{\mathbf{#1}}

\newcommand{\lp}{\left ( }
\newcommand{\rp}{\right ) }

\newcommand{\hc}{\text{H.c.}}

\newcommand{\beq}{\begin{eqnarray*}}
\newcommand{\eeq}{\end{eqnarray*}}

\newcommand{\be}{\begin{eqnarray}}
\newcommand{\ee}{\end{eqnarray}}

\newcommand{\mc}{\mathcal}

\def\lsim{\mathrel{\rlap{\lower4pt\hbox{\hskip1pt$\sim$}}
    \raise1pt\hbox{$<$}}}                

\def\gsim{\mathrel{\rlap{\lower4pt\hbox{\hskip1pt$\sim$}}
    \raise1pt\hbox{$>$}}}                

\begin{document}

\title{Ultracold nonreactive molecules in an optical lattice: \\
connecting chemistry to many-body physics}

\author{Andris Do\c{c}aj}
\thanks{These authors contributed equally.}
\affiliation{Department of Physics and Astronomy, Rice University, Houston, Texas 77005, USA}
\affiliation{Rice Center for Quantum Materials, Rice University, Houston, Texas 77005, USA}
\author{Michael L. Wall} 
\thanks{These authors contributed equally.}
\affiliation{JILA, NIST and University of Colorado, Boulder, Colorado 80309-0440, USA}
\author{Rick Mukherjee}
\affiliation{Department of Physics and Astronomy, Rice University, Houston, Texas 77005, USA}
\affiliation{Rice Center for Quantum Materials, Rice University, Houston, Texas 77005, USA}
\author{Kaden R.~A. Hazzard} \email{kaden.hazzard@gmail.com}
\affiliation{Department of Physics and Astronomy, Rice University, Houston, Texas 77005, USA}
\affiliation{Rice Center for Quantum Materials, Rice University, Houston, Texas 77005, USA}

\begin{abstract}
We derive effective lattice models for ultracold bosonic or fermionic nonreactive molecules (NRMs) in an optical lattice, analogous to the Hubbard model that describes ultracold atoms in a lattice.  In stark contrast to the Hubbard model, which is commonly assumed to accurately describe NRMs, we find that the single on-site interaction parameter $U$ is replaced by a multi-channel interaction, whose properties we elucidate.  The complex, multi-channel collisional physics is unrelated to dipolar interactions, and so occurs even in the absence of an electric field or for homonuclear molecules. We find a crossover between coherent few-channel models and fully incoherent single-channel models as the lattice depth is increased.  We show that the effective model parameters  can be determined in lattice modulation experiments, which consequently measure molecular collision dynamics with a vastly sharper energy resolution  than experiments in an ultracold gas.
\end{abstract}

\maketitle

\emph{Introduction.} The recent production of ultracold ground state molecules opens up far-ranging possibilities for quantum many-body physics. These possibilities stem from 
 molecular properties unavailable to atoms, including strong, long-range electric dipole-dipole interactions and a rich rotational and vibrational structure. Relying on these properties, ultracold molecules can be used for quantum simulation of strongly interacting systems~\cite{carr2009cold,doi:10.1021/cr2003568,wallrole}, quantum information processing~\cite{demilleD2002}, quantum metrology, and exploring chemistry in the quantum regime~\cite{doi:10.1021/cr300092g,0953-4075-39-19-S28,Balakrishnan2001652,PhysRevLett.115.063201,0034-4885-72-8-086401,PhysRevA.90.052716,krems2008cold}. 
The first achieved~\cite{Ni_Ospelkaus_08} and most explored~\cite{Ospelkaus_Peer_08,Ospelkaus_Ni_09,Ospelkaus_Ni_10,PhysRevLett.108.080405,deMiranda2011,Ospelkaus_Ni_10b,PhysRevLett.113.195302,PhysRevLett.109.230403,yan_observation_2013,moses2015creation} ultracold molecule, KRb, undergoes rapid two-body reactions, $\text{KRb}+\text{KRb} \to \text{K}_2+\text{Rb}_2$~\cite{Ospelkaus_Ni_10b}, as do about half of the alkali metal dimers~\cite{PhysRevA.81.060703}. Experiments are underway to cool many of these reactive species~\cite{PhysRevA.86.021602,PhysRevA.89.020702,deiglmayr:formation-LiCs_2008,deiglmayr:permanent-LiCs_2010}.  Even though ultracold reactions offer exciting insights into quantum chemical kinetics and stereodynamics~\cite{deMiranda2011}, they unfortunately limit the cloud lifetime.    
Although reactions are 
 irrelevant for some situations, such as quantum spin models~\cite{wallrole,barnett2006,Gorshkov_Manmana_11,Gorshkov_Manmana_11b,PhysRevA.82.013611,yan_observation_2013}, and can be suppressed in other special cases~\cite{zhu_suppressing_2014,PhysRevLett.105.073202,1367-2630-17-3-035007,1367-2630-17-1-013020}, 
they nevertheless prevent accessing some important physical situations, especially where translational motion of the molecules is important.

\begin{figure}[b]
\includegraphics[width=.97\columnwidth]{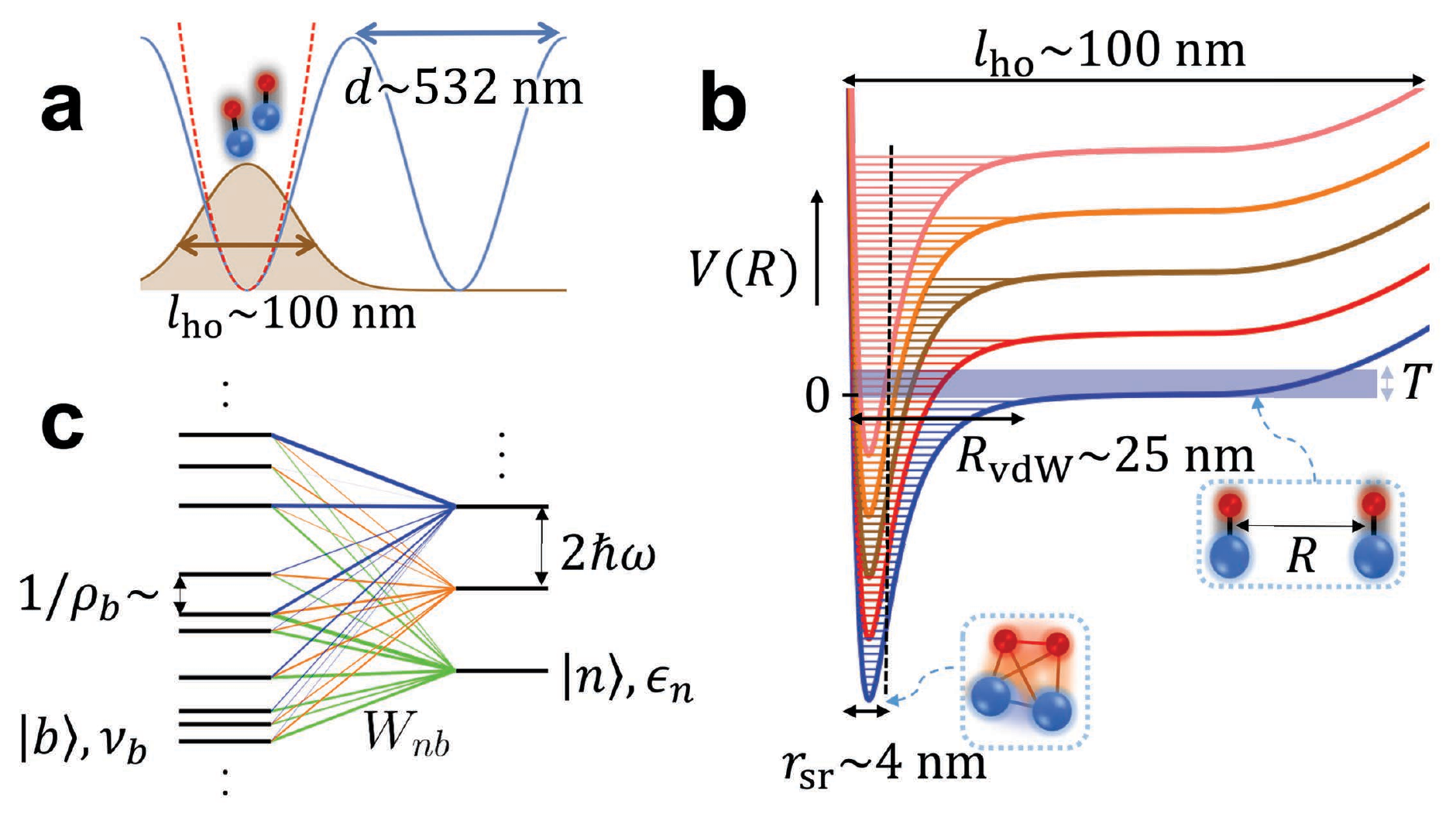}
\caption{(color online) \emph{Nonreactive molecules (NRMs) in an optical trap.} (a) NRMs in a deep optical lattice (solid blue) and its harmonic trap idealization (dashed red).  (b) NRMs, which experience only the harmonic trap at large intermolecular separations $R\gsim l_{\mathrm{ho}}$ are coupled to a dense collection of  collisional complexes at short range $R \lsim r_{\mathrm{sr}}$, which are associated with excited rotational and vibrational interaction channels.  Quoted values are for RbCs~\cite{1367-2630-12-7-073041}. (c) Harmonic oscillator levels $\ket{n}$ with energy $\epsilon_n$ and bound state complexes $\ket{b}$ with energy $\nu_b$ and density of states $\rho_b$ are coupled by state-dependent  $W_{nb}$. \label{fig:NRM-in-lattice}}
\end{figure}
Over the last year experiments have produced ultracold non-reactive molecules (NRMs)~\cite{molony:creation_2014,takekoshi:ultracold-RbCs_2014,park:two-photon_2015,park:ultracold_2015}, and many other experiments are progressing with molecules that are nonreactive or whose reactivity is unknown~\cite{shuman2010laser,1367-2630-17-3-035014,PhysRevLett.110.143001,1367-2630-17-5-055008,PhysRevLett.114.223003,stuhl2012evaporative,zeppenfeld2012sisyphus,chervenkov2014continuous,PhysRevA.92.023404,lemeshko2013manipulation,1367-2630-17-3-035003,KCsNagerl}.  Such molecules are expected to have long lifetimes, but to take advantage of these lifetimes we must understand the molecular interaction properties.  It has recently been argued that the collisions of NRMs are much more complex than for atoms due to an extraordinarily high density of low-energy internal states at short range~\cite{mayle:statistical_2012, mayle:scattering_2013,PhysRevA.89.012714}. This complexity persists in the absence of an electric field, and even for homonuclear molecules~\cite{herbig2003preparation,danzl2008quantum,danzl:ultracold-Cs2_2010,PhysRevLett.109.115303,PhysRevLett.109.115302,frisch2015ultracold,jochim2003bose}. 
Although many interesting scenarios for ongoing and future ultracold molecule experiments involve optical lattices, the complexity of molecular collisions implies that the single-channel pseudopotential approach that is used to derive the  
Hubbard model description of atoms in optical lattices~\cite{jaksch_bruder_98,lewensteinM2007} will rarely apply to NRMs. Therefore, an alternative approach is required to derive a
low-energy effective model that describes NRMs in a lattice.

In this Letter, we provide a framework for deriving effective many-body models for NRMs in deep optical lattices or other tight traps, such as optical tweezers~\cite{Kaufman2012,PhysRevLett.110.133001}. This framework combines transition state theory~\cite{levine:molecular_2010}, random matrix theory (RMT)~\cite{RevModPhys.53.385,mitchell:random_2010}, and  
quantum defect theory~\cite{PhysRevA.26.2441,PhysRevLett.104.113202,PhysRevA.82.020703} and accounts for the separation of short- and long-range scales shown in Fig.~\ref{fig:NRM-in-lattice}.  
We first solve the problem of two NRMs in a single site of an optical lattice 
that are coupled to a dense collection of short-range collision complexes.   Then this on-site solution is coupled to other 
lattice  
sites (via tunneling) to generate a full lattice model.   Our  method resembles those used to derive models of broad two-channel Feshbach resonances~\cite{PhysRevLett.95.243202,PhysRevLett.104.090402,PhysRevLett.109.055302,PhysRevA.87.033601}, even though the physics differs significantly.  
Strikingly, despite the complexity of this system, we show that the model parameters depend universally on only two molecular properties, the van der Waals length $R_{\text{vdW}}$ and the density of bound states  at zero collision energy $\rho_b$. Furthermore, we show that the model parameters can be experimentally characterized with lattice modulation spectroscopy.
  We note that although our quantitative analysis uses a specific collisional model, our general procedure is broadly applicable.

\emph{Formalism.}  Figure~\ref{fig:NRM-in-lattice} shows a schematic of NRMs in a deep optical lattice and our theoretical approach.  Fig.~\ref{fig:NRM-in-lattice}(a) shows how a single site containing two NRMs is approximated  by a harmonic well of frequency $\omega$.  
This approximation is 
quantitatively accurate for on-site properties in a deep lattice~\cite{PhysRevA.77.032726}. 
The harmonic trap quantizes the open channel motion to  harmonic oscillator states that spread out over 
$l_{\mathrm{ho}}\sim 100$nm. 
At short range, $r_{\mathrm{sr}}\sim 4$nm, there are a vast number of internal degrees of freedom,
e.g., vibrations and rotations, each associated with its own interaction potential (channel), each of which may contain several bound states, i.e. \textit{bimolecular collisional complexes}, as illustrated in Fig~\ref{fig:NRM-in-lattice}(b).  
Although at intermolecular separations $R \gg r_{\text{sr}}$ the closed channels are negligible -- since their asymptotic energies are orders of magnitude larger than the temperature -- the bound states couple to the open channel at small $R \lsim r_{\mathrm{sr}}$.
Two NRMs confined to a site of an isotropic 3D optical lattice are described by the multichannel resonance Hamiltonian $H=H_{\text{COM}}+H_{\text{rel}}$, where the center-of-mass (COM) Hamiltonian is  $H_{\text{COM}}=  \omega\left(2n_{\mathrm{COM}}+\ell_{\mathrm{COM}}+3/2\right)$ with $n_{\mathrm{COM}}$ and $\ell_{\mathrm{COM}}$ the COM principal and angular momentum quantum numbers (we set $\hbar=1$ throughout), and the relative coordinate Hamiltonian for the 
$\ell=0$ 
$s$-wave states is 
\be
\label{eq:rel-ham-ex} \hspace{-0.175in} H_{\text{rel}} \!&=&\!\! 
\sum_{n}\! \epsilon_n \!\ket{n}\!\bra{n} \!+\! \sum_b\! \nu_b \!\ket{b}\!\bra{b} 
				\!+\!\sum_{bn}\! \lp W_{nb} \!\ket{n}\!\bra{b}+\hc\rp
\ee 
with $\epsilon_n = \lp2n+3/2\rp \omega$ the energy of harmonic oscillator state $|n\rangle$, $\nu_b$ the energy of bound state $|b\rangle$, and $W_{nb}$ the coupling of harmonic oscillator state $|n\rangle$ to bound state $|b\rangle$~\footnote{We assume trap frequencies are the same for the 
bound states and the molecules  because the polarizabilities and mass are approximately double those of the latter.}.
Fig.~\ref{fig:NRM-in-lattice}(c) displays the structure of this model.  
Although higher partial waves can also contribute in principle, these are suppressed at low energy by Wigner threshold laws~\cite{mayle:scattering_2013}.  
Due to the separation of scales $\{r_{\mathrm{sr}},R_{\text{vdW}}\}\ll l_{\mathrm{ho}}$, we can approximate the bound states as delta functions, in which case the couplings are $W_{nb} = w_bM_n/l_{\text{ho}}^{3/2} $ with $M_n = \sqrt{\Gamma(n+3/2)/\Gamma(n+1)}$, $l_{\text{ho}}=\sqrt{1/(\mu \omega)}$, and the constants $w_b$, which will be determined below, are independent of $n$ and the trap parameters.  Here, $\mu$ is the reduced mass for two molecules and $\Gamma(x)$ is the gamma function.  Actually, naively taking this zero-range coupling limit of Eq.~\eqref{eq:rel-ham-ex} 
leads to divergences. The Supplementary Material describes how  
to regularize this naive prescription to obtain the correct physical limit~\cite{Supp}. 

Although Eq.~\eqref{eq:rel-ham-ex} is an exact description of two NRMs in a harmonic trap when $\{r_{\text{sr}},R_{\text{vdW}}\}\ll l_{\mathrm{ho}}$, 
determining the $\nu_b$ and $w_b$ from an \textit{ab initio} microscopic model of two interacting NRMs is extraordinarily difficult.  Similarly, measuring the parameters is generally infeasible due to the high density of short-range resonant states $\rho_b$; in contrast to simple atoms such as the alkalis, in which  $\rho_b$ is small enough that individual resonances are easily resolved, NRMs are expected to have $\rho_b$ so large that individual resonances are unresolvable at any reasonable temperature~\cite{mayle:statistical_2012,mayle:scattering_2013,PhysRevA.89.012714}.   Hence, the full characterization of the interactions is expected to be beyond the reach of both current theory and experiment.

Instead of an \emph{ab initio} model of the interactions at short range, we use a simple -- yet realistic and potentially accurate --  parameterization to obtain the $\nu_b$ and $w_b$. This interaction model is essentially a Hamiltonian reformulation of the approach introduced in Refs.~\cite{mayle:statistical_2012, mayle:scattering_2013}.   First, we apply random matrix theory (RMT), which is expected to be valid in the often-relevant case where molecular collisions are chaotic~\cite{PhysRevA.89.020701,maier2015broad,maier2015emergence,PhysRevA.92.020702,PhysRevA.88.052701,PhysRevA.89.012714,Fryeetal}. Specifically, the 
$\nu_b$ are the eigenvalues of matrices sampled from the Gaussian orthogonal ensemble (GOE). 
The GOE  probability distribution is $P_H(H_b) = {\mc B} e^{-\operatorname{Tr} H_b^2/2\sigma^2}$ over $H_b$ in the set of $N_b \times N_b$ real symmetric matrices, ${\mc B}$ is a normalization factor, and  
$\sigma=\sqrt{N_b}/(\pi\rho_b)$ is chosen to match the molecule's $\rho_b$ for 
$N_b\rightarrow\infty$.  The couplings $w_b$ are chosen from the probability distribution $P_w(w_b) = {\mc C} e^{-w_b^2/2\sigma_w^2} $
where ${\mc C}$ is a normalization constant and 
the standard deviation $\sigma_w$ depends on the molecule.  
In particular, we relate $\sigma_w$ to molecular parameters by matching the decay rate of a bound state calculated in the free space $\omega\to 0 $ limit of our theory to the physical decay rate. The former is obtained by Fermi's golden rule. The latter is approximated by combining Rice-Ramsperger-Kassel-Marcus (RRKM) transition-state theory to account for the decay rate at $R\sim r_{\mathrm{sr}}$ with quantum defect theory (QDT) to  describe propagation in the long-range van der Waals tail of the intermolecular potential~\cite{mayle:statistical_2012, mayle:scattering_2013,ruzic:quantum_2013}.  
RRKM is a standard chemical approximation whose core assumption is that the molecules' configurations are in equilibrium until they cross a reaction surface, which then is never re-crossed~\cite{levine:molecular_2010}. 
QDT is an exact treatment of the potential tail that is crucial to obtain the Wigner threshold laws.  This procedure, explained in more detail in the Supplementary Material~\cite{Supp}, yields $\sigma_w =  \sqrt{{8 R_{\text{vdW}}}/({\pi^3\mu \rho_b})}\Gamma(3/4)$; 
remarkably, $\sigma_w$ depends only on $R_{\mathrm{vdW}}$ and  $\rho_b$.
 
\begin{figure}[t]
\includegraphics[width=.97\columnwidth,angle=0]{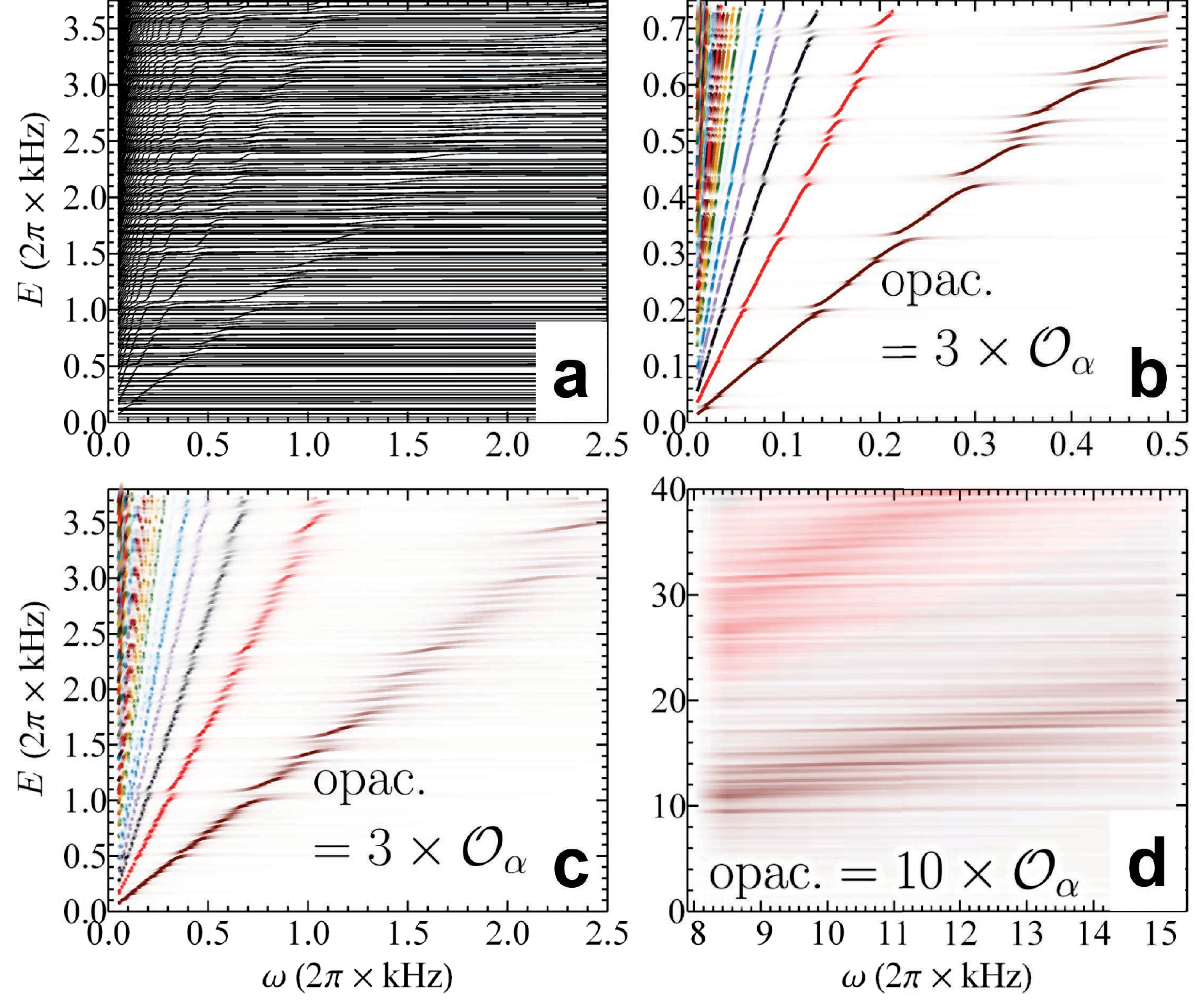}
\caption{(color online) \emph{States of two NRMs in 
a harmonic trap.}  
(a) Eigenvalues versus trap frequency $\omega$ for RbCs. Bound states $\ket{b}$ (roughly horizontal) hybridize with harmonic oscillator states $|n\rangle$ (roughly diagonal). 
(b-d) By setting the opacity of a point to the associated eigenstate's open-channel probability $\mathcal{O}_{\alpha}$, the states relevant to the lattice model emerge from panel (a)'s tangle of lines.
(b) Weak trap, isolated resonances.
(c) Intermediate lattice depths, rich structure of resonances  coupled to harmonic oscillator states. 
(d) Deep trap, universal dissipative limit: resonances merge into a broad near-continuum. 
All of these regimes are experimentally relevant even though for RbCs in an isotropic 3D lattice the isolated resonance limit occurs at $\omega$ for which the harmonic oscillator approximation breaks down.
 \label{fig:spectrum}}
\end{figure}

 \emph{Spectrum of two NRMs in a lattice site.}    We numerically solve~Eq.~\eqref{eq:rel-ham-ex} using the $\nu_b$ and $W_{nb}$ described above 
 to obtain the eigenvalues $E_\alpha$ and eigenstates $\ket{\alpha}$~\cite{Supp}.  
 Fig.~\ref{fig:spectrum} displays the behavior of $E_\alpha$ and $\ket{\alpha}$ using exemplary parameters corresponding to RbCs~\cite{1367-2630-12-7-073041,mayle:statistical_2012},  $\rho_b\approx 1/\mbox{nK}\approx (2\pi \times 20 \text{Hz})^{-1}$ and $R_{\mathrm{vdW}}\approx 25$nm, though we stress that our model applies to any NRM.   Fig.~\ref{fig:spectrum}(a) shows the spectrum for $H_{\text{rel}}$ as a function of harmonic oscillator frequency $\omega$, neglecting the COM energy common to all states.  Harmonic oscillator energies  increase with increasing $\omega$, while short-range bound state energies are independent of $\omega$.
For a given $\omega$, most of these eigenstates can be ignored since only those with non-negligible fraction on the open channel will be accessed experimentally.
Therefore, the results are more informative if one sets the opacity of a point associated with $\ket{\alpha}$ to its weight on the open-channel (trap states) $\mathcal{O}_{\alpha}\equiv \sum_{N,M} \left|O_{\alpha;N,M}\right|^2$, where $O_{\alpha;N,M}\equiv \braket{\alpha|N,M}$ is the overlap of eigenstate $\alpha$ with the two-particle open-channel state $|N,M\rangle$ 
labeled by 
the first and second particle's harmonic oscillator states, $N$ and $M$, respectively.   Fig.~\ref{fig:spectrum}(b) displays the eigenstates for $\omega \lesssim 2\pi\times 500$Hz weighted 
 in this fashion, showing that most bound states are uncoupled from the trap states in this (small-$\omega$) \emph{isolated resonance} frequency regime. 
Here NRMs are described by single- or few-channel models, just like atoms. In contrast to the continuum where the  spread in energy is set by the temperature $k_BT$, and where, since $k_B T\gg \rho_b^{-1}$, many collisional complexes are coupled, the trap states' energies are precisely quantized and can couple significantly only to a single collisional complex.   
Fig.~\ref{fig:spectrum}(d) displays the spectrum in deep traps $\omega\sim 2\pi\times 15$kHz, which corresponds to trap depths similar to those used in common optical lattice experiments.  
The $W_{nb}$ are larger for the larger $\omega$ -- simply because they 
are proportional to the probability amplitude for two molecules to have zero separation -- and so 
couple a broader energy range of collisional complexes. 
Because
  $W_{nb}\ll  \omega$, there is still little mixing between open channel states for this $\omega$.  However, many overlapping resonances  couple to each $\ket{n}$, resulting in a smeared near-continuum of levels which we call the (large-$\omega$) \emph{universal dissipative limit}, for reasons clarified later.  
Fig.~\ref{fig:spectrum}(c) shows the open-channel weighted spectrum for intermediate trap depths, $\omega\sim 2\pi\times 2$kHz.  In many ways this is the most novel regime, with 
a rich structure of non-isolated, but not completely overlapping, resonances.

We expect that all of these regimes are experimentally accessible. 
 For RbCs, the universal dissipative limit and some of the intermediate regime occur  where the harmonic oscillator approximation (and single-band, tight binding approximation for the lattice model presented below) will typically be valid, $\omega \gsim 5$kHz. In contrast, we present the isolated resonance limit mainly for its relevance to other NRMs.
For NRMs with a smaller $\rho_b$, as expected for lighter NRMs (larger rotational constants) such as NaK~\cite{park:two-photon_2015,park:ultracold_2015}, 
 or smaller $R_{\text{vdW}}$, as predicted for a range of molecules in 
 Ref.~\cite{julienne:universal_2011},  the crossover will occur at larger, more accessible, $\omega$. In particular, the crossover occurs  when $W_{nb}/\rho_b\sim 1$; from our earlier expressions the scaling of the crossover is  then $  (\omega \rho)^{3/4}  \sqrt{R_{\text{vdW}}(\mu/\rho_b)^{1/2}}\sim 1$.

\begin{figure}
\includegraphics[width=.97\columnwidth,angle=0]{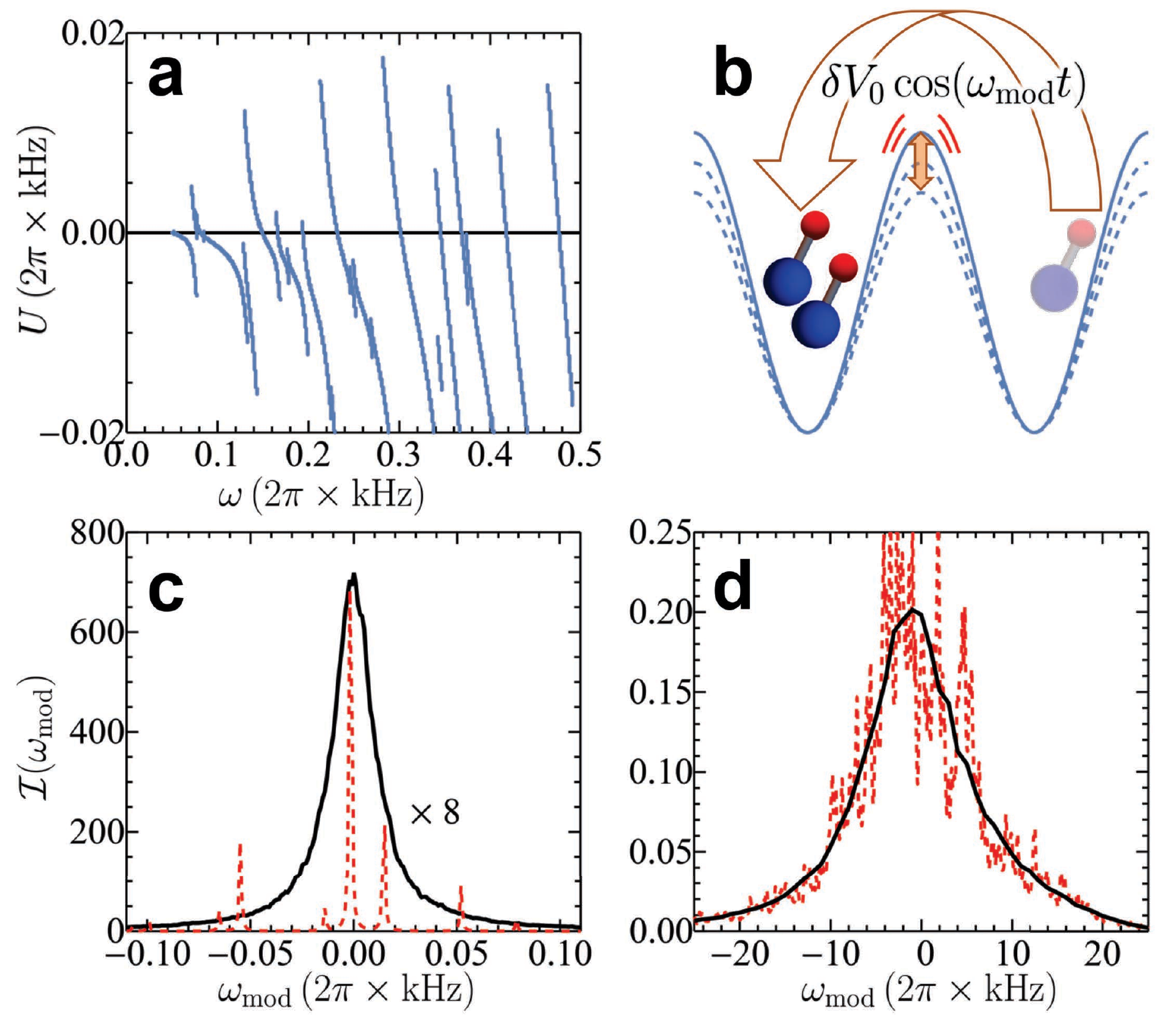}
\caption{
\textit{Effective lattice model parameters and experimental characterization.} (a) Interaction energies $U_\alpha$ versus $\omega$
for 
$\alpha$s that have weight on the trap ground state $\vert O_{\alpha;0,0}\vert^2>0.2$. 
(b) Lattice modulation spectrum ${\mc I}(\omega_{\text{mod}})$: one slightly modifies the lattice depth at frequency $\omega_{\text{mod}}$ and measures the induced tunneling, for example the rate of change of double occupancies. 
This probes the lattice model parameters. (c)  
${\mc I}(\omega_{\text{mod}})$ for $\omega=2\pi \times 250\text{Hz}$ for one realization of the random parameters (dashed line), and averaged over 10,000 realizations (solid line).  (d) Same as (c), but for $\omega= 2\pi\times 15\text{kHz}$, in the universal dissipative   limit, and with the solid line averaging 50 realizations. Unlike the isolated resonance limit, ${\mc I}(\omega_{\text{mod}})$ ``self-averages": one realization approaches the averaged spectrum and is thus independent of molecular details.
\label{fig:extract-U}
}
\end{figure}

 \emph{Full lattice model.} Knowing the two-particle single-site solution, we couple sites to determine the effective lattice model $H_{\mathrm{latt}}$ valid when at most two molecules per site are relevant.  For bosons 
(adapting to fermions or additional open channels is straightforward),  
\be
\label{eq:Lattmodel} 
\hspace{-0.15in}H_{\mathrm{latt}} &=& 
-t\sum_{\langle i,j \rangle} c^\dagger_i c_j + \sum_i \lp 
	\sum_{\alpha}U_\alpha n_{i,\alpha} + \lp 3/2\rp  \omega n_{i}
	 \rp\!. 
\ee
This Hamiltonian harbors two new features compared to the usual Bose-Hubbard model. (i) There are multiple interaction channels $\alpha$
with interaction energy 
$U_\alpha= E_\alpha-3\omega/2$. (ii) $c_i^\dagger$ and $c_i$ are modified from the usual creation/annihilation operators: $c^\dagger_i \ket{\text{vac}}=\ket{0}_i$, $c^\dagger_i \ket{0}_i=\sum_\alpha \sqrt{2} O_{\alpha;0,0} \ket{\alpha}_i$, and $c^{\dagger}_i|\alpha\rangle_i=0$  where $\ket{0}_i$ is the site-$i$ single-particle ground state, i.e. $\ket{N=0}_i$. 
We have defined $n_{i,\alpha}=\ket{\alpha}_i\!\bra{\alpha}_i$ and $n_i=|0\rangle_i\langle 0|_i+2\sum_{\alpha}  n_{i,\alpha}$.
Eq.~\eqref{eq:Lattmodel} is a many-channel generalization of Refs.~\cite{Duan_05,von_Stecher_Gurarie_11}.  
We have ignored the collisional complexes' tunneling $t_{\text{cc}}\ll t$ because they have approximately twice the molecules' polarizability and mass, and also ignored other terms that scale as $t_{\text{cc}}$, such as molecules and complexes exchanging sites~\cite{von_Stecher_Gurarie_11}.
Figure~\ref{fig:extract-U}(a) shows the on-site interactions $U_\alpha$ as a function of $\omega$. For each $\omega$, we plot the $U_\alpha$ for $\ket{\alpha}$ that have 
${\mc O}_\alpha>0.2$.
For these states and the plotted range of $\omega$, typically one or two channels are relevant. 

\emph{Lattice modulation spectroscopy.} The interaction and overlap parameters $U_{\alpha}$ and $O_{\alpha;0,0}$ appearing in the lattice model Eq.~\eqref{eq:Lattmodel} can be characterized, both theoretically and experimentally, via lattice modulation spectroscopy~\cite{PhysRevA.74.041604,jordens2008mott,PhysRevLett.102.065301} as illustrated in Fig.~\ref{fig:extract-U}(b).  In this procedure, the depth of the lattice potential is modulated periodically with frequency $\omega_{\mathrm{mod}}$, which induces tunneling from a non-interacting configuration where molecules are on neighboring lattice sites into an interacting configuration with two molecules on the same site.  For weak modulation the response is given by linear response theory and we find a spectrum proportional to
\be
{\mc I}(\omega_{\text{mod}}) &=& \sum_\alpha \frac{|O_{\alpha;0,0}|^2}{\omega_{\text{mod}}-U_\alpha+i 0^+}\, ,
\ee
where $0^{+}$ is a positive infinitesimal.  
Figs.~\ref{fig:extract-U}(c) and~(d) show ${\mc I}(\omega_{\text{mod}})$ for different trap frequencies $\omega$. 
The dashed red curves are for a single realization of the RMT parameters, while the solid black curves average over many realizations.  
In the isolated resonance regime (Fig.~\ref{fig:extract-U}(c)), well-separated peaks appear for each eigenstate $\ket{\alpha}$ with amplitude determined by $\ket{\alpha}$'s weight on the ground harmonic oscillator state, $|O_{\alpha;0,0}|^2$,  and the physics is well-described by a few-channel model.  For sufficiently small $\omega$, a single peak would dominate, and a one-channel model describes the physics.  In this limit, the exact locations of the resonances are random, but can be predicted statistically. 
In the universal dissipative limit (Fig.~\ref{fig:extract-U}(d), large $\omega$), many resonances smear together to give a continuous curve. For sufficiently large $\omega$ (but still small enough that band-mixing is negligible), the spectrum for a single realization approaches the average over many realizations. Moreover, this spectrum approaches a Lorentzian with width $\gamma$; in fact, Eq.~\eqref{eq:Lattmodel} reduces to a single channel model again, but with  imaginary $U=-i\gamma$.
In between these limits is the most novel regime, where several channels contribute but not so many that they can be described by an incoherent single-channel model. 
In closing we stress that 
Eq.~\eqref{eq:Lattmodel} holds more generally than our specific collision model, and the spectra ${\mc I}(\omega_{\text{mod}})$ give a direct way of characterizing the lattice model parameters.

\emph{Conclusion.} We have put forth a framework for deriving effective lattice models for NRMs in optical potentials.
We elucidated the models' structure  and how to probe their parameters experimentally.
We found a crossover between a coherent one-channel model 
(conventionally used to describe atoms) for weak traps  to an entirely incoherent Markovian single-channel model (often used to describe reactive molecules) for deep traps.  In between, novel multi-channel models emerge, whose many-body physics is an exciting frontier to explore.

Our results also show that optical lattice experiments can probe chemical properties: the bimolecular complexes. Unlike thermal gas measurements, which probe 
a range of energies set by the temperature,
molecules in a trap have a tunable, exactly quantized energy.
Thus NRMs in a lattice or optical tweezer array~\cite{Kaufman_Lester_14,lester:rapid_2015}  provide a high-energy-resolution ``chemical complex microscope" to probe the complexes' energies and couplings.

Clear next steps are to explore more elaborate collisional models and ways to control the interactions, e.g. via electric fields~\cite{PhysRevA.85.022703}. 
Finally, we note that our methods apply to other systems, such as lanthanide atoms~\cite{maier2015broad,maier2015emergence} whose $\rho_b$ lies between molecules and alkali atoms.

\begin{acknowledgements} 
We thank John Bohn, Brandon Ruzic, Svetlana Kotochigova, Michael Foss-Feig, Ana Maria Rey, Daniel Sheehy, Shah Alam, and Ian White for useful discussions.  
KRAH acknowledges the Aspen Center for Physics for its hospitality while part of this work was performed. This work was supported with funds 
from the Welch foundation, Grant No. C-1872.  MLW acknowledges support from the NRC postdoctoral fellowship program. 
\end{acknowledgements}

\bibliography{NRM-1site-v4}

\clearpage

\onecolumngrid

\renewcommand{\theequation}{S\arabic{equation}}
\setcounter{equation}{0}

\renewcommand{\thefigure}{S\arabic{figure}}
\setcounter{figure}{0}

\renewcommand\thesection{}

\hrulefill
\begin{center}
\Large\textbf{Supplemental Material for ``Ultracold nonreactive molecules in an optical lattice: connecting chemistry to many-body physics''}
\end{center}

\section{Microscopic derivation of Eq.~(\ref{eq:rel-ham-ex}) and couplings $W_{nb}$} 

In this section, we derive Eq.~\eqref{eq:rel-ham-ex} in the main text and the couplings $W_{nb}$ obtained there. 
The exact microscopic Hamiltonian $\hat{H}$ for a pair of molecules contains the kinetic energy of the nuclei and electrons along with the Coulomb interactions amongst them. $\hat{H}$ is re-written without approximation in the basis of BO  adiabatic states by defining the intermolecular separation $\v{R}$ as the ``slow variable." 
Although calculating these states and the Hamiltonian projected in these coordinates remains intractable, we use this expansion only to define a basis, and we do not need to compute the states. 

We can write an arbitrary wavefunction as a sum of the open $\ket{O}$ and closed $\ket{B}$ channel components as $\ket{\psi(R)} = \psi_O(R)\ket{O}  + \sum_B \psi_B(R)\ket{B}$, where the kets specify the channel and the wavefunctions $\psi_B(R)$ and $\psi_O(R)$ contain the associated spatial dependence.  The diagonal part of the Hamiltonian is sketched in  Fig.~\ref{fig:NRM-in-lattice}(b); furthermore, the Hamiltonian connects states $\ket{b}=\psi_b(R)\ket{B}$ and $\ket{n}=\psi_n(R)\ket{O}$ via the non-adiabatic couplings
\be
W_{nb}=\braket{n|\hat{H}|b} &=& \int \!  dR\, \psi_n(R) {\mc W}(R) \psi_b(R)
\ee
where ${\mc W}(R)\equiv\braket{O|\hat{H}|B}$, $\psi_b(R)$ is the relative coordinate bound state wavefunction for state $\ket{b}$, and $\psi_n(R)={\mc N}_n e^{-R^2/2l_{\text{ho}}^2}L_n^{(1/2)}((R/l_{\text{ho}})^2)$ is the $n^{\mathrm{th}}$ $s$-wave harmonic oscillator eigenfunction. Here $L_n^{(a)}$ is the generalized Laguerre polynomial and ${\mc N}_n=l_{\text{ho}}^{-3/2} \sqrt{\Gamma(n+1)/\Gamma(n+3/2)}$.  For a bound state which is tightly bound ($\psi_b(R) \propto \delta(R)$),  the matrix element is $W_{nb} = \psi_n(0) A_b$  for some constant $A_b$ that is independent of $n$ and the trap properties.    Using $L^{(1/2)}_n(0)=(2/\sqrt{\pi})\Gamma(n+3/2)/\Gamma(n+1)$, we obtain
\be 
W_{nb}&=&w_bM_n/l_{\text{ho}}^{3/2}\, ,\label{eq:zero-range-coupling}
\ee
where $M_n=\sqrt{\Gamma(n+3/2)/\Gamma(n+1)}$ and $w_b$ is an unknown constant (related to $A_b$), in the naive zero-range limit.

\section{Regularizing the zero-range limit, Eq.~(\ref{eq:rel-ham-ex})} 

Here we describe how the naive zero-range limit of Eq.~\eqref{eq:rel-ham-ex} and Eq.~\eqref{eq:zero-range-coupling} may be regularized to obtain the physical Hamiltonian. The naive zero-range limit approximates the bound state wavefunctions as having zero range with fixed energies $\nu_b$  when $W_{nb}=0$. However, the true physical limit is a bit more subtle: the \textit{physical} bound state energies are indeed some finite, fixed set of numbers, but these are not the same as the $\nu_b$ in Eq.~\eqref{eq:rel-ham-ex}. Rather, the physical energies correspond to the eigenenergies after coupling to the continuum, which gives a divergent shift of the eigenenergies away from the $\nu_b$. Although the regularization of the one- and two-channel models is standard~\cite{pethick02} and requires only a (diverging) shift in the bare bound state energies, the regularization of a multi-channel model such as ours requires new couplings, and to our knowledge has not appeared previously in the literature. 

We define  $\Lambda$ as the energy cutoff for the open channel, which cuts off the sum over harmonic oscillator states $n$ such that $\epsilon_n<\Lambda$; explicitly, the sum runs to $n^*=\operatorname{Floor}[\Lambda/(2\omega)-3/4]$. We find that the Hamiltonian that properly accounts for the physical zero-range limit is
\be
H_{\text{rel}}(\Lambda) &=& \sum_{n \text{ with } \epsilon_n<\Lambda} \epsilon_n \ket{n}\!\bra{n} +\sum_{b,b'} \lp -\delta_{bb'} \nu_b^* + \sqrt{\frac{\mu ^3\Lambda}{2}} w_b w_{b'}\rp \ket{b}\!\bra{b'} 
+\sum_{b,n \text{ with } \epsilon_n<\Lambda} \lp \frac{w_{b} M_n}{l_{\text{ho}}^{3/2}} \ket{n}\!\bra{b}+\hc\rp  \label{eq:H-rel-reg}
\ee
for $\Lambda\rightarrow \infty$ [in which case it suffices to take $n^*=\Lambda/(2\omega)$]. The key addition to Eq.~\eqref{eq:rel-ham-ex} to obtain the physical zero-range limit is the term proportional to $\sqrt{\Lambda}$ that couples bound states $\ket{b}$ and $\ket{b'}$ and shifts the energy of each bound state $\ket{b}$.  
It can be readily verified that (I) the physical properties of Eq.~\eqref{eq:H-rel-reg} are independent of $\Lambda$ for $\Lambda/\omega \gg 1$ [i.e. Eq.~\eqref{eq:H-rel-reg} is a regularization of Eq.~\eqref{eq:rel-ham-ex}] and (II) Eq.~\eqref{eq:H-rel-reg} reproduces the low-energy properties of the true microscopic physical Hamiltonian (i.e. it is the appropriate physical regularization). These statements follow from performing second order non-degenerate perturbation theory (i.e. a Schrieffer-Wolff transform) to obtain the effective Hamiltonian for Eq.~\eqref{eq:rel-ham-ex} that holds in the truncated Hilbert space with some open channel cutoff, and showing that this effective Hamiltonian agrees with Eq.~\eqref{eq:H-rel-reg} and is independent of $\Lambda$ as $\Lambda\to\infty$.

\section{Combining transition state theory with quantum defect theory}

Although random matrix theory gives the structure of the bound state-open channel couplings and bound state energies, it does not determine the overall scale ($w_b$) for the couplings. In the main text, we determine the standard deviation of the $w_b$'s by combining transition state theory (TST) with quantum defect theory (QDT). Here we elaborate this calculation, which is equivalent to that of Refs.~\cite{mayle:statistical_2012, mayle:scattering_2013}, but implemented in a slightly different way. To obtain $w_b$, the essential idea (as described in the main text)  is to equate the dissociation rate $\gamma$ of a bound state (collisional complex) into the continuum of two independent molecules as determined in the $\omega\rightarrow 0$ limit of Eq.~\eqref{eq:rel-ham-ex} to the rate predicted by TST~+~QDT ($\gamma_{\rm TST+QDT}$). Applying Fermi's Golden rule to Eq.~\eqref{eq:rel-ham-ex}, the dissociation rate $\gamma$ is  
\be
\hspace{0.5in}
\gamma &=& 2\pi \frac{w_b^2}{l_{\text{h.o.}}^3}\sum_n M_n^2 \delta(\nu_b - \epsilon_n) \nonumber \\
&=& \frac{\pi \mu^{3/2} \sqrt{\nu_b} w_b^2}{\sqrt{2}} \hspace{1.2in} \text{for $\omega\to 0$} \label{eq:cont-diss}
\ee
The second line follows for $\omega\to 0$ by replacing the sum with an integral and expanding the $\Gamma$ functions appearing in $M_n=\sqrt{\Gamma(n+3/2)/\Gamma(n+1)}$ for large values of their argument. 

On the other hand, the rate determined by TST+QDT is determined as the product of two factors, $\gamma_{\rm TST+QDT}= {\mc A}(\nu_b) \gamma_{\text{TST}}$. For a given complex, $\gamma_{\text{TST}}=2/(\pi \rho_b)$ is the TST approximation to the decay rate from a bound state to the open channel at short-range ($r_{\text{sr}}$ in Fig.~\ref{fig:NRM-in-lattice}), for a barrier-less reaction $(\text{complex}) \to (\text{molecule})+(\text{molecule})$ with a single open channel~\cite{levine:molecular_2010,mayle:statistical_2012, mayle:scattering_2013}. The ${\mc A}(\nu_b)$ obtained from QDT is the probability of the two molecules to propagate from this short range regime to outside the van der Waals potential ($R_{\text{vdW}}$ in Fig.~\ref{fig:NRM-in-lattice}). 
TST is a standard and often reasonable approximation in chemistry~\cite{levine:molecular_2010}. TST's core assumption is that a molecule's configurations are in equilibrium until it crosses a reaction surface, after which it never re-crosses the surface. Traditionally, it also assumes the motion is classical, which is valid for short-range distances $R\lsim r_{\text{sr}}$ where the van der Waals energy is large, leading to a large kinetic energy and thus effectively classical dynamics. In contrast, for larger $R$, the quantum effects are crucial: the quantum propagation in the van der Waals potential for the ultracold systems of interest qualitatively alters the scattering, giving rise to Wigner threshold laws. This propagation can be accounted for by a factor ${\mc A}(E)=\frac{2^{3/2}\sqrt{\mu} }{\pi}R_{\text{vdW}}\Gamma(3/4)^2\sqrt{E}$ that is calculated from QDT~\cite{mayle:scattering_2013,mayle:statistical_2012,ruzic:quantum_2013}, where $E$ is the scattering energy. Thus, the total dissociation rate from TST~+~QDT for the bound state complexes is
\be
\gamma_{\rm TST+QDT} &=&  \frac{2^{5/2}\sqrt{\mu} }{\pi^2\rho_b}R_{\text{vdW}}\Gamma(3/4)^2\sqrt{\nu_b}.  
\label{eq:TST-QDT-diss}
\ee 
We determine the $w_b$ by matching the dissociation rate $\gamma$ given by Eq.~\eqref{eq:cont-diss} to that expected 
from the combination of TST~+~QDT, Eq.~\eqref{eq:TST-QDT-diss}. Actually, the rate in our model Eq.~\eqref{eq:cont-diss} depends on $b$, since $w_b$ is sampled randomly for each bound state, so we can only match these two expressions on average, determining $\overline{w_b^2}$. Since $\overline{w_b}=0$, we have $\sigma_w^2=\overline{w_b^2}$, and the matching gives
\be
\label{eq:RMT+TST} \sigma_w &=& \sqrt{\frac{8 R_{\text{vdW}}}{\pi^3 \mu\rho}} \Gamma(3/4)\, ,
\ee
which is the expression quoted and used in the main text. 

\section{Numerical implementation and convergence}

\begin{figure}[h]
	\includegraphics[width=.7\columnwidth]{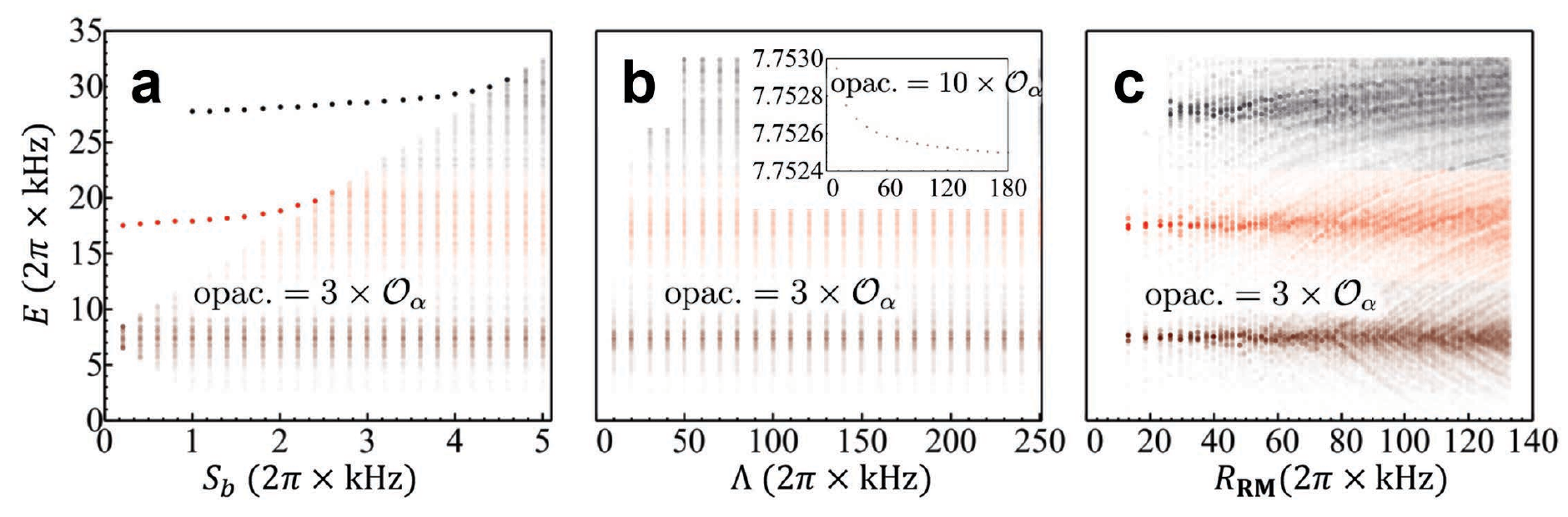}
	\caption{(color online) \emph{Convergence of numerics.} 
		Convergence for a $\omega=2\pi \times 5$~kHz trap, where eigenvalues are plotted with weights as in Figs.~2(b-d) in the main text. (a) Convergence with respect to selected energy window of bound states $S_b$.  
		(b) Convergence with respect to energy cutoff $\Lambda$ on harmonic oscillator states (open channels).   
		(c) Statistical convergence with respect to dimension $N_b$ of random matrix from which the bound state energies $\nu_b$ are sampled, parameterized by the radius $R_{\text{RM}}$ of the resulting eigenvalue distribution. 
		\label{fig:convergence}}
\end{figure}

Our numerical procedure follows three steps: (I) Generate the $\nu_b$ in Eq.~\eqref{eq:rel-ham-ex} from RMT. (II) Generate the couplings $w_b$ according to RMT together with TST+QDT using the probability distributions in the main text with variance given by Eq.~\eqref{eq:RMT+TST}. (III) Diagonalize $H_{\text{rel}}$, with appropriate regularization.  All of our results -- e.g. the spectra presented in the text -- then follow from straightforward processing of the eigenvalues and eigenvectors.   This section demonstrates the convergence of the numerical methods in these steps, summarized in Fig.~\ref{fig:convergence}.  To obtain the $\nu_b$, one calculates the eigenvalues of $N_b \times N_b$ symmetric random matrices sampled according to the Gaussian orthogonal ensembles (GOE), as described in the text. The GOE is obtained in the $N_b \rightarrow \infty$ limit, and the average distribution of eigenvalues have a density of states peaked at zero energy with energy width $R_{\text{RM}}=2\sqrt{N_b}\sigma$. One wishes $N_b$ to be large enough so that the eigenvalue distribution becomes sufficiently broad to have a nearly constant density of states that matches the density of complexes $\rho_b$ of the molecule being considered, over the energy range $\delta E$ that is coupled to the states of interest. For example, for the harmonic oscillator state in Figs.~3(c-d), $\delta E$ is roughly the width of the spectrum.  

As a naive method, one could simply use the $\nu_b$ generated in this manner in Eq.~\eqref{eq:rel-ham-ex}. However, this becomes prohibitively expensive even for modest lattice depths: the Hamiltonian is a dense $O(N_b) \times O(N_b)$ matrix, requiring $O(N_b^3)$ computational time to diagonalize. Already for $\omega\sim 15$kHz the bound states within a $\delta E \sim 20$kHz window around a harmonic oscillator couple to it, and roughly a thousand states lie in this window. Note that in order to obtain a density of states that is constant to within $\sim 10\%$ over this window may then require $N_b \gsim 10,000$. Thus, it can be beneficial to select only $\nu_b$ in a window of width $S_b$ much less than the total width of the density of states but larger than $\delta E$ for use in diagonalizing Eq.~\eqref{eq:rel-ham-ex}. This can reduce the number of bound states by an order of magnitude.   However, there is now an additional convergence parameter $S_b$ to keep track of: $S_b$ must be large enough that all relevant bound states to the physics of interest are included, while simultaneously $N_b$ is large enough that the density of states of $\nu_b$ is constant over the physically relevant $\delta E$. 

Figure~\ref{fig:convergence} shows  an example of the convergence of the numerics for $\omega = 2\pi \times 5$kHz with respect to all of our convergence parameters: $S_b$, $\Lambda$, and $N_b$ as parameterized by the width of the GOE $R_{\text{RM}}=2\sqrt{N_b}\sigma$, in panels (a-c), respectively. Considering the eigenstates near the ground harmonic oscillator state, the spectrum is clearly well-converged for a bound state window $S_b\sim 2\pi \times 2$kHz and for an energy cut off of $\Lambda \sim 50$kHz. With respect to the bound state density matrix width, the system has converged when $R_{\text{RM}}\sim 80$kHz. Here, the apparent ``stripes" in the eigenvalues occur because the random noise is correlated for nearby $R_{\text{RM}}$ as we choose increasingly large random matrices in this convergence plot by first generating a large matrix and then plotting the eigenvalues of increasingly large submatrices for increasing $R_{\text{RM}}$.

\end{document}